\documentclass[12pt]{iopart}

\usepackage{cite}

\begin{document}

\title{More integrals of products of Airy functions}
\author{Francisco M. Fern\'{a}ndez}

\address{INIFTA (UNLP, CCT La Plata-CONICET), Divisi\'on Qu\'imica Te\'orica,
Blvd. 113 S/N,  Sucursal 4, Casilla de Correo 16,
1900 La Plata, Argentina}\ead{fernande@quimica.unlp.edu.ar}

\maketitle

\begin{abstract}
By means of a modified hypervirial theorem we derive simple expressions for
the integrals of products of Airy functions. Present results contain earlier
ones as particular cases.
\end{abstract}

Airy functions appear in many problems of mathematical physics. For that
reason some time ago Gordon\cite{G69} and Albright\cite{A77} derived
integrals of products of Airy functions that do not appear in commonly
available tables of integrals and functions. However, they are necessary for
several problems of physical interest\cite{G69,A77}.

The purpose of this letter is to propose a simple approach for the
derivation of such integrals. It is based on a straightforward modification
of the well known hypervirial theorems\cite{FC87} and may also be useful for
other similar problems.

Let $A(x)$ and $B(x)$ be solutions to the eigenvalue equations
\begin{eqnarray}
\hat{L}A &=&aA,\;\hat{L}B=bB  \nonumber \\
\hat{L} &=&\hat{D}^{2}-x,\;\hat{D}=\frac{d}{dx}  \label{eq:Airy}
\end{eqnarray}
We can easily prove that for any linear operator $\hat{O}$
\begin{equation}
\int A[\hat{L},\hat{O}]B\,dx=(a-b)\int A\hat{O}B\,dx+W(A,\hat{O}B)
\label{eq:HVT_O}
\end{equation}
where $[\hat{L},\hat{O}]=\hat{L}\hat{O}-\hat{O}\hat{L}$ is the commutator, $%
W(u,v)=uv^{\prime }-u^{\prime }v$ is the Wronskian and the prime indicates
differentiation with respect to $x$.

We need the following two cases
\begin{equation}
\lbrack \hat{L},f\hat{D}]=f^{\prime \prime }\hat{D}+2f^{\prime }(\hat{L}+x)+f
\label{eq:[L,fD]}
\end{equation}
and
\begin{equation}
\lbrack \hat{L},g]=g^{\prime \prime }+2g^{\prime }\hat{D}  \label{eq:[L,g]}
\end{equation}
where $f$ and $g$ are differentiable functions of $x$. It follows from
equation (\ref{eq:HVT_O}) with $\hat{O}=f\hat{D}$ and $\hat{O}=g$ that
\begin{eqnarray}
\int Af^{\prime \prime }\hat{D}B\,dx&+&2b\int Af^{\prime }B\,dx+2\int
Axf^{\prime }B\,dx+\int AfB\,dx =  \label{HVT_fD} \\
&&(a-b)\int Af\hat{D}B\,dx+W(A,fB^{\prime })
\end{eqnarray}
and
\begin{equation}
\int Ag^{\prime }\hat{D}B\,dx=\frac{a-b}{2}\int AgB\,dx-\frac{1}{2}\int
Ag^{\prime \prime }B\,dx+\frac{1}{2}W(A,gB)  \label{eq:HVT_g}
\end{equation}
respectively. Notice that we can use the latter equation twice with $%
g=f^{\prime }$ and with $g^{\prime }=f$ to remove the two integrals in the
former equation that contain $\hat{D}B=B^{\prime }$. Thus we derive the
following expression
\begin{eqnarray}
\int AfB\,dx &+&2\int Axf^{\prime }B\,dx-\frac{1}{2}\int Af^{\prime \prime
\prime }B\,dx+(a+b)\int Af^{\prime }B\,dx  \nonumber \\
&-&\frac{(a-b)^{2}}{2}\int AFB\,dx=W(A,fB^{\prime })-\frac{1}{2}%
W(A,f^{\prime }B)  \nonumber \\
&+&\frac{a-b}{2}W(A,FB)  \label{eq:master_1}
\end{eqnarray}
where $F=\int f\,dx$. From this master equation we easily obtain any
integral of the form $\int x^{n}A(x)B(x)\,dx$ such as the particular cases
considered by Gordon\cite{G69}. Notice that he chose functions
of the form $A[\alpha (R+\beta _{1})]$ but we can set the scale $\alpha =1$
without loss of generality so that $\beta _{1}=a$ and $\beta _{2}=b$ in our
notation. For example, if $F=1$ we obtain his equation (B.10):
\begin{equation}
\int A(x)B(x)\,dx=\frac{1}{b-a}[A^{\prime }(x)B(x)-A(x)B^{\prime }(x)]
\label{eq:Gordon1}
\end{equation}

If we choose $a=b=0$ then the master equation (\ref{eq:master_1}) gives us
all the integrals of the form $I_{n}(y_{1},y_{2})=$ $\int
x^{n}y_{1}y_{2}\,dx $ considered by Albright\cite{A77}. For example, when $%
f=x^{n}$ we obtain the recurrence relation
\begin{eqnarray}
\int x^{n}A(x)B(x)\,dx &=&\frac{n(n-1)(n-2)}{2(2n+1)}\int
x^{n-3}A(x)B(x)\,dx+\frac{x^{n+1}A(x)B(x)}{2n+1}  \nonumber \\
&&-\frac{n(n-1)x^{n-2}A(x)B(x)}{2(2n+1)}+\frac{nx^{n-1}[A(x)B^{\prime
}(x)+A^{\prime }(x)B(x)]}{2(2n+1)}  \nonumber \\
&&-\frac{x^{n}A^{\prime }(x)B^{\prime }(x)}{2n+1},\; n=0,1,\dots  \label{eq:x^nAB}
\end{eqnarray}
On the other hand, equation (\ref{eq:HVT_g}) enables us to connect the
integrals $I_{n}(y_{1},y_{2}^{\prime })$ with the $I_{m}(y_{1},y_{2})$ ones.
Finally, if we take into account that
\begin{equation}
\int fy_{1}^{\prime }y_{2}^{\prime }\,dx=fy_{1}^{\prime }y_{2}-\int
f^{\prime }y_{1}^{\prime }y_{2}\,dx-\int xfy_{1}y_{2}\,dx
\end{equation}
we easily express the integrals $I_{k}(y_{1}^{\prime },y_{2}^{\prime })$ in
terms of those discussed above.

In our opinion the modified hypervirial theorems provide a most convenient
simple and systematic
approach to the calculation of integrals of products of Airy functions.

\end{document}